\documentclass[a4paper,11pt]{article}
\usepackage{jheppub} 
\usepackage{lineno}
\usepackage{booktabs}

\arxivnumber{xxxx} 

\title{\boldmath Neural-Network Holographic Model of the QCD Phase Transition under Lattice and HRG Constraints}

\author[a]{Dexing Zhu,}
\author[a]{Liqiang Zhu,}
\author[b,c]{Xun Chen,}
\author[a]{Defu Hou}
\author[d,e]{and Kai Zhou}
\affiliation[a]{Institute of Particle Physics and Key Laboratory of Quark and Lepton Physics (MOE), Central China Normal University, \\Wuhan 430079, China}
\affiliation[b]{School of Nuclear Science and Technology, University of South China, \\Hengyang 421001, China}
\affiliation[c]{INFN --- Istituto Nazionale di Fisica Nucleare,  Sezione di Bari, \\Via Orabona 4, 70125, Bari, Italy}
\affiliation[d]{School of Science and Engineering, The Chinese University of Hong Kong, \\Shenzhen (CUHK-Shenzhen), Guangdong, 518172, China}
\affiliation[e]{School of Artificial Intelligence, The Chinese University of Hong Kong, \\Shenzhen (CUHK-Shenzhen), Guangdong, 518172, China}

\emailAdd{chenxun@usc.edu.cn}
\emailAdd{houdf@ccnu.edu.cn}
\emailAdd{zhoukai@cuhk.edu.cn}

\abstract{Within a neural-network-based holographic framework, we incorporate lattice QCD (LQCD) and Hadron Resonance Gas (HRG) data to train the model and predict the location of the QCD critical endpoint (CEP).
The training dataset consists of the entropy density, baryon number susceptibility, and baryon density.
The metric warp factor $A(z)$ and the gauge kinetic function $f(z)$ are parameterized by neural networks and determined through the training procedure.
The resulting model reproduces the equation of state at vanishing chemical potential in good agreement with both LQCD and HRG data.
Extending the analysis to finite chemical potential, we solve the equations of motion and obtain thermodynamic observables consistent with LQCD results at finite density.
After incorporating the HRG constraints, the predicted position of the CEP shifts toward larger chemical potentials compared to recent studies.
We further employ symbolic regression to derive analytic expressions for $A(z)$ and $f(z)$, providing convenient functional forms for future phenomenological applications.
Finally, we perform a data-driven validation using synthetic thermodynamic data generated from an existing analytical holographic model.
The neural-network framework reproduces the corresponding CEP location with good accuracy, showing close agreement within numerical uncertainties.}

\begin{document}
\maketitle

\section{Introduction}\label{sec.I}

Ultra-relativistic heavy-ion collisions performed at the Relativistic Heavy Ion Collider (RHIC) and the Large Hadron Collider (LHC) have provided compelling experimental evidence for the creation of a new state of strongly interacting matter known as the Quark--Gluon Plasma (QGP)~\cite{ALICE:2012wz, CMS:2013bza, ALICE:2015ylo, CMS:2016itk, ALICE:2017fvd, CMS:2017uoy, CMS:2020gji, ALICE:2021nwy, Muller:2012zq, Busza:2018rrf,PHENIX:2004vcz,STAR:2010vob,Chen:2024aom,Wang:2022fwq,Wang:2021xpv,Wang:2019vhg,Xu:2018fog,Jin:2018lbk,Liu:2017rjm,Zhu:2025edv,Li:2024uzk,Xie:2024xbn,He:2023zin,Chen:2024eaq,Xu:2016fns}. Understanding the properties of this strongly interacting QCD matter naturally leads to the investigation of the QCD phase diagram, which remains one of the central frontiers in modern nuclear physics. 
The transition from hadronic matter to the QGP is governed by non-perturbative dynamics that challenge traditional theoretical methods~\cite{Guenther:2020jwe,Aarts:2023vsf,HotQCD:2014kol}.
While perturbative QCD provides accurate predictions at high energy scales, the low-energy regime---characterized by color confinement and chiral symmetry breaking---requires non-perturbative tools.
Lattice QCD (LQCD) has successfully established that the transition at vanishing baryon chemical potential ($\mu_B = 0$) is a smooth crossover~\cite{Aoki:2006we, Borsanyi:2013hza, Adam:2025phc}.
However, extending these first-principles calculations to finite baryon density is hindered by the numerical sign problem~\cite{Philipsen:2012nu, Aarts:2015tyj, Nagata:2021ugx, deForcrand:2009zkb}, leaving the high-density regime and the potential existence of a Critical Endpoint (CEP) largely unexplored.

AdS/CFT~\cite{Maldacena:1997re, Witten:1998qj, Gubser:1998bc, Aharony:1999ti}, offers a powerful complementary approach. Holography maps strongly coupled gauge theories to weakly coupled gravitational theories in a higher-dimensional spacetime. 
This mapping enables the calculation of quantities -- such as thermodynamic observables and transport coefficients -- in regimes inaccessible to LQCD~\cite{Rougemont:2023gfz, Grefa:2021qvt, Rougemont:2015wca}. 
Bottom-up holographic models, such as the Einstein-Maxwell-Dilaton (EMD) framework, have been particularly successful in phenomenologically reproducing QCD properties~\cite{Gursoy:2007er, Gursoy:2008bu, DeWolfe:2010he, Zhao:2023gur,Khan:2026uqc}.
A central challenge in bottom-up holography is the \textit{inverse problem}: determining the precise functional forms of the bulk metric and background fields that correspond to physical QCD data~\cite{Grefa:2021qvt, Ecker:2025vnb}.
Traditional approaches rely on parameterized analytical ans\"atze, such as the Soft-wall model~\cite{Karch:2006pv}.

Recently, machine learning (ML) techniques have been widely applied in high-energy physics~\cite{Ma:2023zfj, Zhou:2023pti, He:2023zin, Wang:2023kcg, Pang:2024kid, Li:2025tqr, Shi:2026jkp}, emerging as powerful tools for tackling complex inverse problems and non-perturbative dynamics. 
In particular, ML provides a robust data-driven framework for extracting hidden structures from high-dimensional datasets and for constructing effective models beyond traditional analytical approaches. 
In the context of holographic QCD, several works have attempted to build data-driven holographic models by incorporating LQCD results, experimental measurements, or phenomenological constraints into neural-network architectures. 
These studies span a broad range of applications, including meson spectroscopy \cite{Hashimoto:2021ihd, Chen:2025kqb, Zhang:2026zoz, ThomasArun:2025uyi}, the QCD phase transition and critical phenomena~\cite{Zeng:2025tcz, Cai:2024eqa, Deng:2026aht}, shear viscosity in strongly coupled plasma~\cite{Yan:2020wcd}, chiral condensate reconstruction~\cite{Hashimoto:2018bnb}, and the heavy-quark potential~\cite{Mansouri:2024uwc, Luo:2024iwf}. 
Additional related developments can be found in~\cite{Hashimoto:2018ftp, Ahn:2024jkk, Ahn:2024gjf,Song:2020agw, Zhu:2025xiz, Jeong:2025omu}.

In our previous studies, holographic models were constructed through gradient-based parameter optimization constrained by LQCD data~\cite{Chen:2024ckb,Chen:2024mmd,Chen:2025goz}. 
Since the Hadron Resonance Gas (HRG) model~\cite{Andronic:2017pug} provides a reliable non-perturbative description of QCD thermodynamics in the confined phase, particularly for $T \lesssim T_c$, it serves as a natural infrared constraint for the equation of state (EoS). 
Motivated by this complementarity between low- and high-temperature regimes, we develop a data-driven holographic framework that systematically incorporates high-temperature LQCD results together with low-temperature HRG constraints, allowing for a consistent reconstruction of the EoS throughout the $(T,\mu_B)$ plane. 
By employing Physics-Informed Neural Network (PINN) to parameterize the bulk metric and gauge kinetic coupling, we achieve a precise description of the crossover transition without relying on rigid analytical ans\"atze. Our model predicts the existence of a CEP at $(T_c, \mu_c) = (0.089, 0.922)$ GeV, connecting the smooth crossover at zero density to a first-order phase transition at high density.  
Additionally, we utilize symbolic regression to derive analytic expressions for the reconstructed background functions, offering a transparent holographic dual for QCD matter.  
The paper is organized as follows: Section~\ref{sec.II} outlines the EMD formalism and the neural network training methodology. Section~\ref{sec.III} presents the numerical results, including the thermodynamic variables, the phase diagram, and the symbolic regression analysis. Finally, a summary is provided in Section~\ref{sec.IV}.


\section{Theoretical Descriptions}\label{sec.II}

\subsection{Einstein-Maxwell-Dilaton Framework}

We employ a bottom-up five-dimensional EMD model to construct the holographic dual of the QCD EoS~\cite{Gursoy:2007cb,Li:2011hp,He:2013qq,Yang:2014bqa,Chen:2024ckb,Yang:2015aia,Dudal:2017max,Dudal:2018hae,Chen:2018msc,Chen:2020ath,Zhou:2020xqi,Chen:2020pty,Chen:2025hdi,Zhu:2025gma,Zeng:2025tcz,Zhu:2025gxo,Chen:2025fpd}. The action in the Einstein frame is given by:
\begin{equation}
    S=\frac{1}{16\pi G_5}\int d^5x\,\sqrt{-g}\bigg[
R-\frac{f(\phi)}{4}F^2-\frac12\,\partial_\mu\phi \partial^\mu\phi - V(\phi)
\bigg],
\label{eq:action}
\end{equation}
where $G_5$ is the five-dimensional Newton constant. $\phi$ is the dilaton field, which is used to break the conformal symmetry. $V(\phi)$ is the potential, and $f(\phi)$ is the gauge kinetic function coupling the Maxwell field $A_\mu$ to the background geometry. 
We treat the gauge coupling $f(\phi)$ as a dynamical function of the holographic coordinate, to be determined via data-driven reconstruction.

The metric ansatz is chosen as:
\begin{equation}
ds^2=\frac{L^2 e^{2A(z)}}{z^2}\left[-g(z)\,dt^2+\frac{dz^2}{g(z)}+d\vec{x}^{\,2}\right],
\label{eq:metric}
\end{equation}
where $z$ is the holographic coordinate ($z=0$ is the boundary), and $L=1$ is the AdS radius. The function $A(z)$ is the warp factor. The blackening function $g(z)$ determines the horizon location $z_h$ via $g(z_h)=0$.

\subsection{Neural Network Reconstruction}

To faithfully reproduce the LQCD EoS without relying on ad-hoc analytical ansätze, we employ the PINN approach. 
We utilize two independent neural networks to reconstruct the background geometry and the gauge sector:
\begin{enumerate}
    \item Metric Network: The warp factor $A(z)$ is parameterized as:
    \begin{equation}
        A(z) = z^2 \cdot \mathcal{N}_A(z; \theta_A),
    \end{equation}
    where $\mathcal{N}_A$ is a neural network and the factor $z^2$ ensures the UV boundary condition $A(0)=0$ is strictly satisfied.
    \item Gauge Network: The gauge kinetic function $f(z)$ which governs the coupling between the background geometry and the baryon density, is parameterized by a second independent neural network:
    \begin{equation}
        f(z) = \mathcal{N}_f(z; \theta_f).
    \end{equation}
\end{enumerate}
This allows the model to learn the non-trivial density dependence implicitly contained in the lattice baryon susceptibility $\chi_2^B$, without restricting the solution space to a specific analytical class.
Once $A(z)$ and $f(z)$ are fixed by the networks, the remaining background functions $g(z)$ and $A_t(z)$ are derived by solving the Einstein-Maxwell equations: 
\begin{equation}
\begin{aligned}
&\phi^{\prime}(z)=\sqrt{6\left(A^{\prime 2}-A^{\prime \prime}-2 A^{\prime} / z\right)}, \\
&g(z)=1+\frac{1}{\int_0^{z_h} y^3 e^{-3 A} d y}\bigg[-\int_{0}^z y^3 e^{-3 A} d y+\left(\frac{\mu}{\int_0^{z_h} \frac{y}{e^{A}f} d y}\right)^2 \hat{G} \bigg],\\
&A_t(z)=\mu \frac{\int_{z_h}^z \frac{y}{e^Af} d y}{\int_{z_h}^0 \frac{y}{e^A f} d y},\\
&V(z)=-3 z^2 g e^{-2 A}\bigg[A^{\prime \prime}+A^{\prime}\left(3 A^{\prime}-\frac{6}{z}+\frac{3 g^{\prime}}{2 g}\right)-\frac{1}{z}\left(-\frac{4}{z}+\frac{3 g^{\prime}}{2 g}\right)+\frac{g^{\prime \prime}}{6 g}\bigg],
\end{aligned}
\end{equation}
where $\mu$ is the chemical potential.
The expression of $\hat{G}$ is
\begin{equation}
    \hat{G}=\left|\begin{array}{ll}
\int_0^{z_h} y^3 e^{-3 A} d y & \int_0^{z_h} y^3 e^{-3 A} d y \int_{0}^y \frac{x}{e^{A}f} d x \\
\int_{z_h}^z y^3 e^{-3 A} d y & \int_{z_h}^z y^3 e^{-3 A} d y \int_{0}^y \frac{x}{e^{A}f} d x
\end{array}\right|.
\end{equation}

\subsection{Thermodynamics and Conserved Quantities}

The thermodynamic observables of the boundary system are established through the holographic dictionary. 
Consequently, the temperature $T$, entropy density $s$, and baryon number density $\rho$ evaluate to:
\begin{equation}
    \begin{aligned}
    T=&\frac{1}{4 \pi} \frac{z_h^3 e^{-3 A\left(z_h\right)}}{\int_0^{z_h} y^3 e^{-3 A(z)} d y}\times \bigg[1-\left(\frac{\mu}{\int_0^{z_h} \frac{y}{e^{A }f} d y}\right)^2 \\
      &\qquad \times \left(\int_0^{z_h} y^3 e^{-3 A} d y \cdot \int_0^{z_h} \frac{x}{e^A f} d x-\int_0^{z_h} y^3 e^{-3 A} d y \int_0^y \frac{x}{e^A f} d x\right)\bigg],\\
    s=&\frac{e^{3 A\left(z_h\right)}}{4 G_5 z_h^3},\\
    \rho =& \frac{1}{16 \pi G_5} \frac{\mu}{\int_{0}^{z_h} \frac{y}{e^A f}dy}.
    \label{}
\end{aligned}
\end{equation}
Given the $s$ and $\rho$, the free energy $F$ and pressure $P$ follow from integration of the appropriate thermodynamic identity~\cite{Critelli:2017oub, Chen:2024mmd}.
\begin{equation}
dF = -dP = -s\,dT-\rho\,d\mu.
\label{eq:firstlaw}
\end{equation}
The $F$ is normalized to zero at $(T, \mu) = (0, 0)$. With this convention, the energy density can be expressed as
\begin{equation}
\epsilon = -P + sT + \mu\rho.
\label{eq:epsilon}
\end{equation}
We also consider the trace anomaly 
\begin{equation}
I=\epsilon-3P.
\label{eq:trace}
\end{equation}
The specific heat at constant volume is computed from the entropy~\cite{DeWolfe:2010he},
\begin{equation}
C_V = T\left(\frac{\partial s}{\partial T}\right)_\mu.
\label{eq:Cv}
\end{equation}
The squared speed of sound for systems with non-zero chemical potential is obtained via~\cite{Yang:2017oer, Li:2020hau, Gursoy:2017wzz}
\begin{equation}
C_s^2=
\frac{s}{
T\left(\frac{\partial s}{\partial T}\right)_\mu
+\mu\left(\frac{\partial \rho}{\partial T}\right)_\mu}.
\label{eq:cs2}
\end{equation}
The second-order baryon number susceptibility is deﬁned as
\begin{equation}
\chi_2^B = \frac{1}{T^2}\frac{\partial \rho}{\partial \mu}.
\label{eq:chi2}
\end{equation}
The thermodynamic consistency of our Neural-Network Holographic Model (NNHM) is rigorously examined through the behavior of these derived quantities.

\subsection{Training Strategy}
The NNHM are optimized by minimizing a total loss function $\mathcal{L}$ that quantifies the deviation from LQCD data. The training process integrates data from both zero and finite chemical potentials:
\begin{enumerate}
    \item Metric Training ($\mu=0$): $\mathcal{N}_A$ is trained using entropy density $s(T)$ at vanishing chemical potential.
    \item Gauge Training ($\mu \neq 0$): $\mathcal{N}_f$ is trained using baryon number susceptibility $\chi_2^B(T)$ at $\mu = 0$ and baryon density $\rho(T, \mu)$ at $\mu_B / T = 1$.
\end{enumerate}
To address the scarcity of lattice data in the low-temperature regime, we augment the training dataset with HRG model predictions for $T < 0.13$ GeV. This hybrid dataset ensures that the neural network learns the correct hadronic degrees of freedom, thereby determining more precise thermodynamic properties. For this study, $G_5$ = 4.17069 was obtained through neural network training.
\footnote{The code can be found at: \href{https://github.com/De-Xing-Zhu/2-1_flavor_network}{https://github.com/De-Xing-Zhu/}}

\section{Results and discussion}\label{sec.III}

A fundamental challenge in bottom-up holographic QCD is the construction of a bulk geometry that simultaneously captures the physics of the deconfined QGP and the confined hadronic phase within a unified framework. 
Traditional models often rely on hand-crafted analytical ansätze for the warp factor $A(z)$ or dilaton potential $V(\phi)$, which inherently introduces human-imposed bias and often fails to reproduce the hadronic limit at low temperatures ($T < 0.13$ GeV). 

In this work, we implement a data-to-holography pipeline that solves the inverse problem by treating $A(z)$ and the gauge-dilaton coupling $f(z)$ as deep-learning-optimized functional profiles. 
To anchor the model across the entire phase space, we adopt a hybrid training strategy:
(2+1)--flavor LQCD data~\cite{HotQCD:2014kol,Bellwied:2015lba} is used for the high-temperature regime.
For the low-temperature region, where lattice calculations encounter computational bottlenecks, the model is constrained by predictions from the HRG model, these predictions are generated using the Thermal-FIST package~\cite{Vovchenko:2019pjl}, a C++ based software tool specifically designed for studying heavy-ion collisions and hadronic equations of state.

\subsection{Thermodynamic Validation at $\mu_B = 0$}

\begin{figure*}[htbp]
    \centering
    \begin{minipage}{0.45\textwidth}
        \centering
        \includegraphics[width=\textwidth]{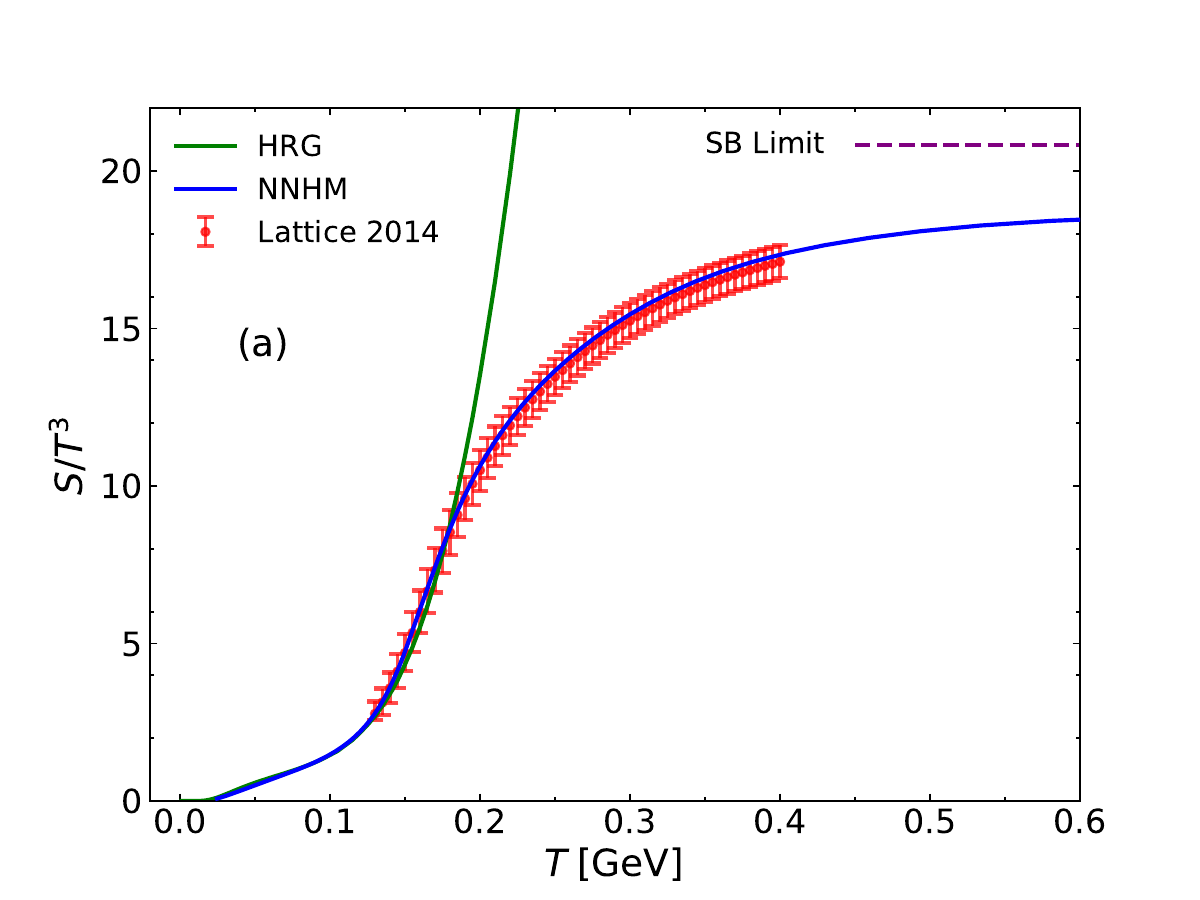}
    \end{minipage}
    \hspace{0.1cm}
    \begin{minipage}{0.45\textwidth}
        \centering
        \includegraphics[width=\textwidth]{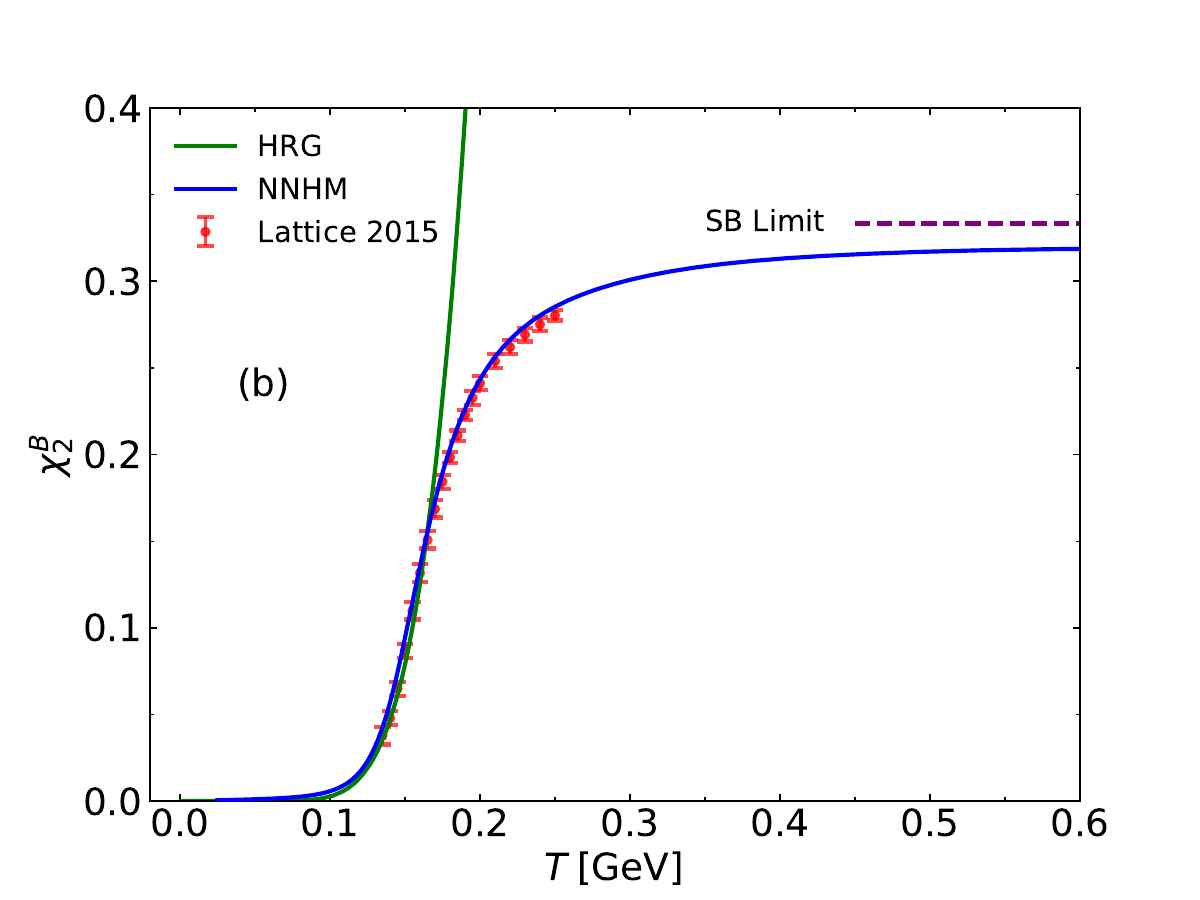}
    \end{minipage}
    \vspace{-0.2cm}
    \begin{minipage}{0.45\textwidth}
        \centering
        \includegraphics[width=\textwidth]{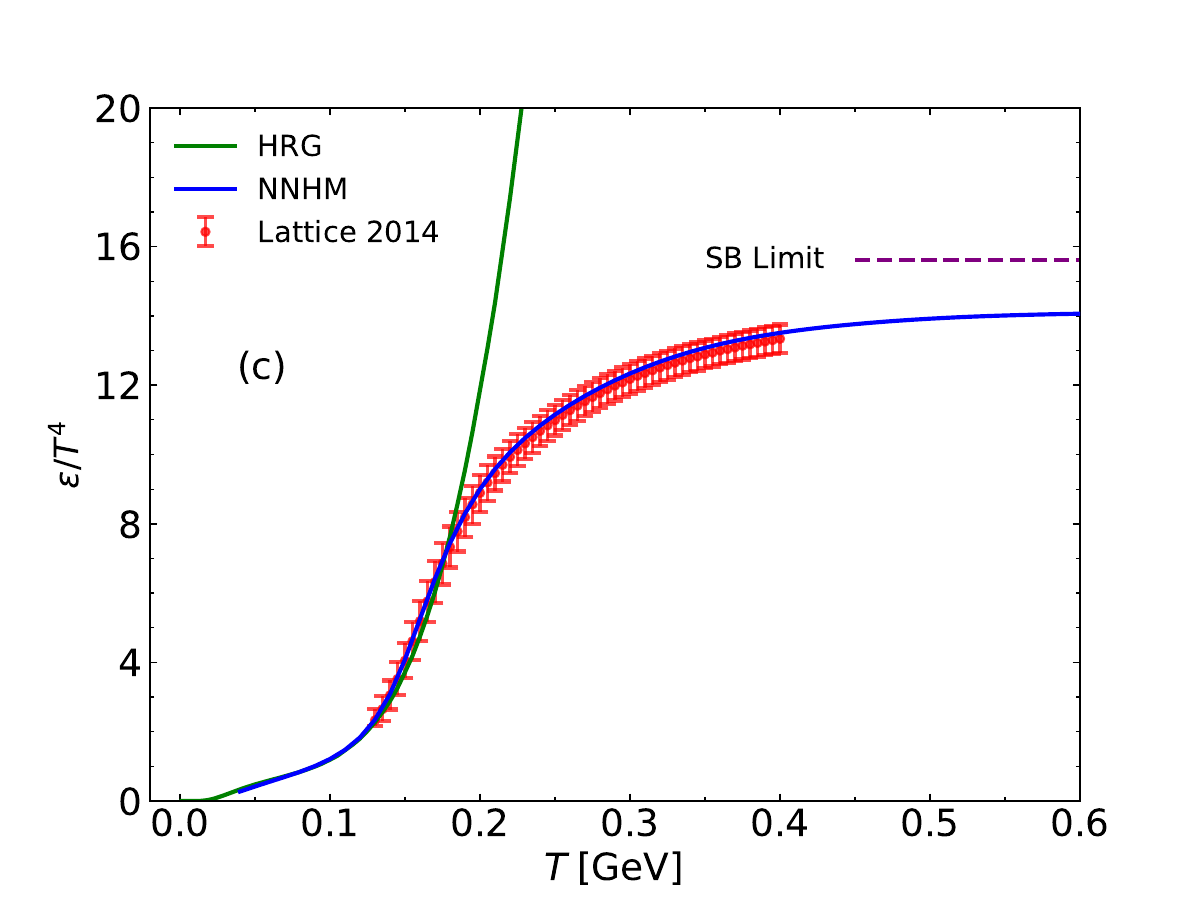}
    \end{minipage}%
    \hspace{0.1cm}%
    \begin{minipage}{0.45\textwidth}
        \centering
        \includegraphics[width=\textwidth]{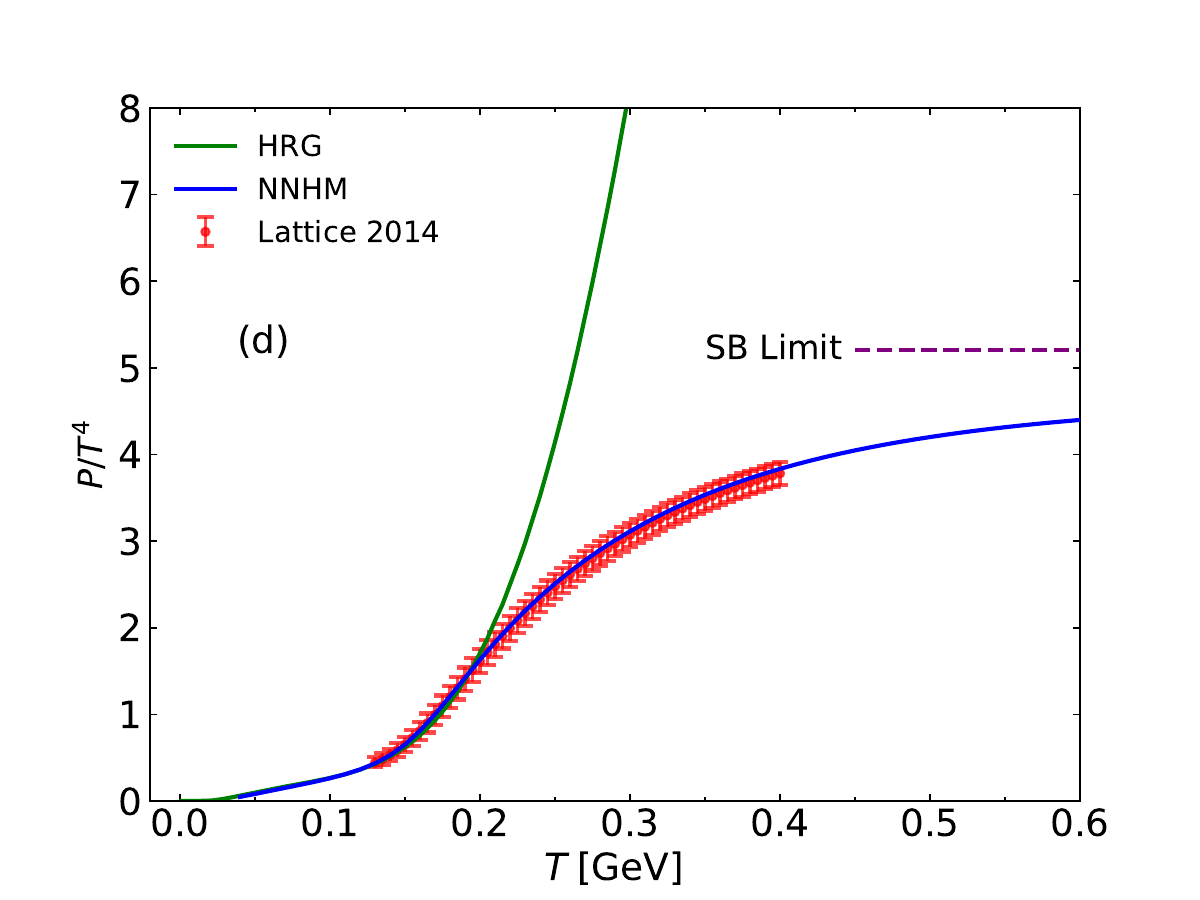}
    \end{minipage}
    \vspace{-0.2cm} 
    \begin{minipage}{0.45\textwidth}
        \centering
        \includegraphics[width=\textwidth]{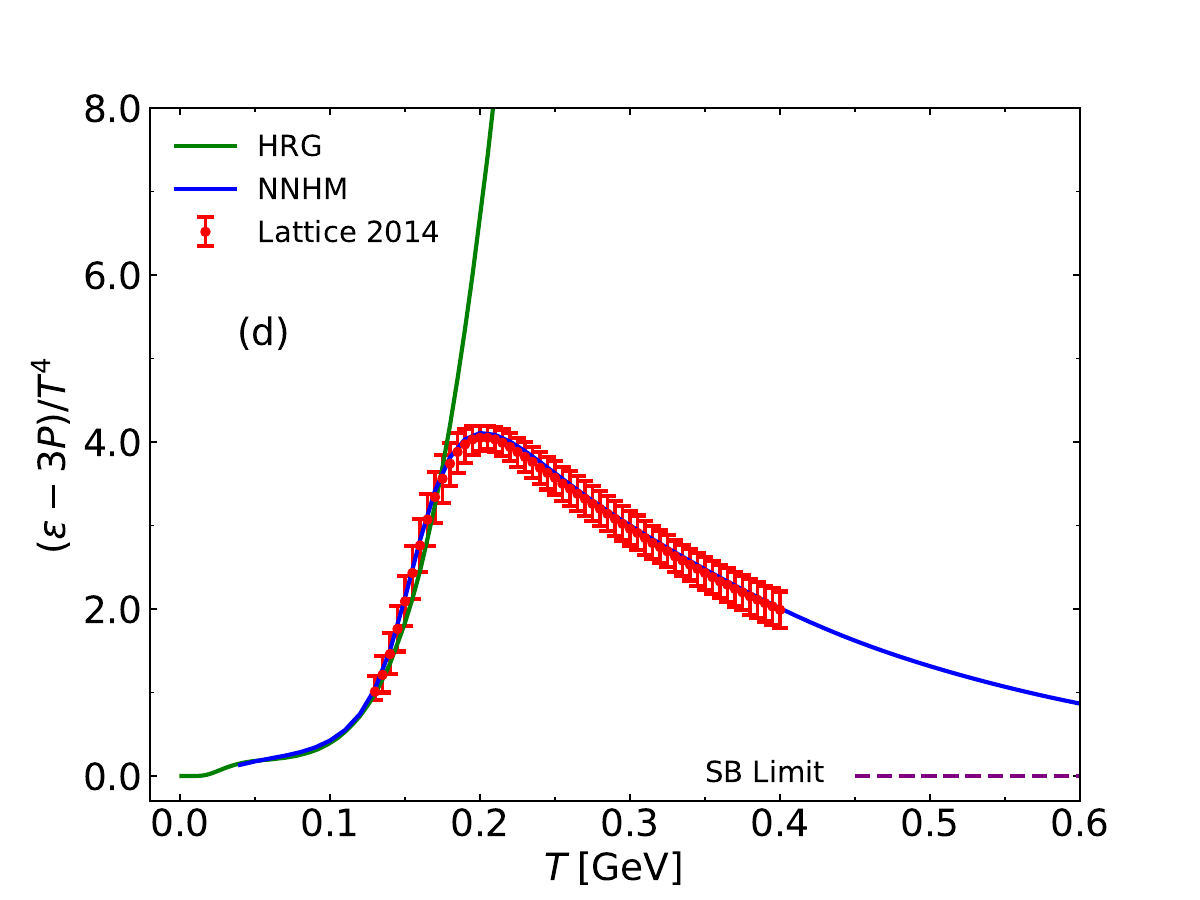}
    \end{minipage}
    \hspace{0.1cm}
    \begin{minipage}{0.45\textwidth}
        \centering
        \includegraphics[width=\textwidth]{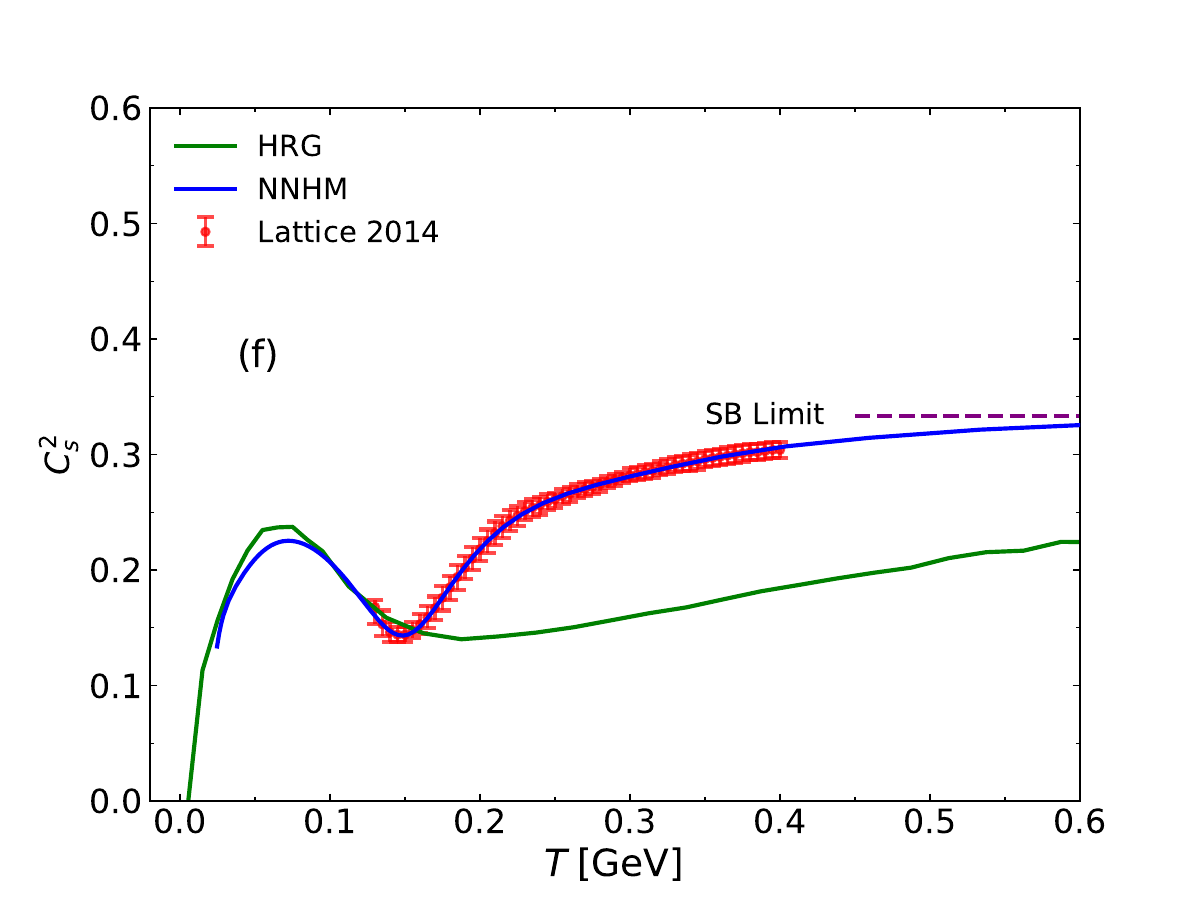}
    \end{minipage}
    \vspace{-0.2cm}
    \begin{minipage}{0.45\textwidth}
        \centering
        \includegraphics[width=\textwidth]{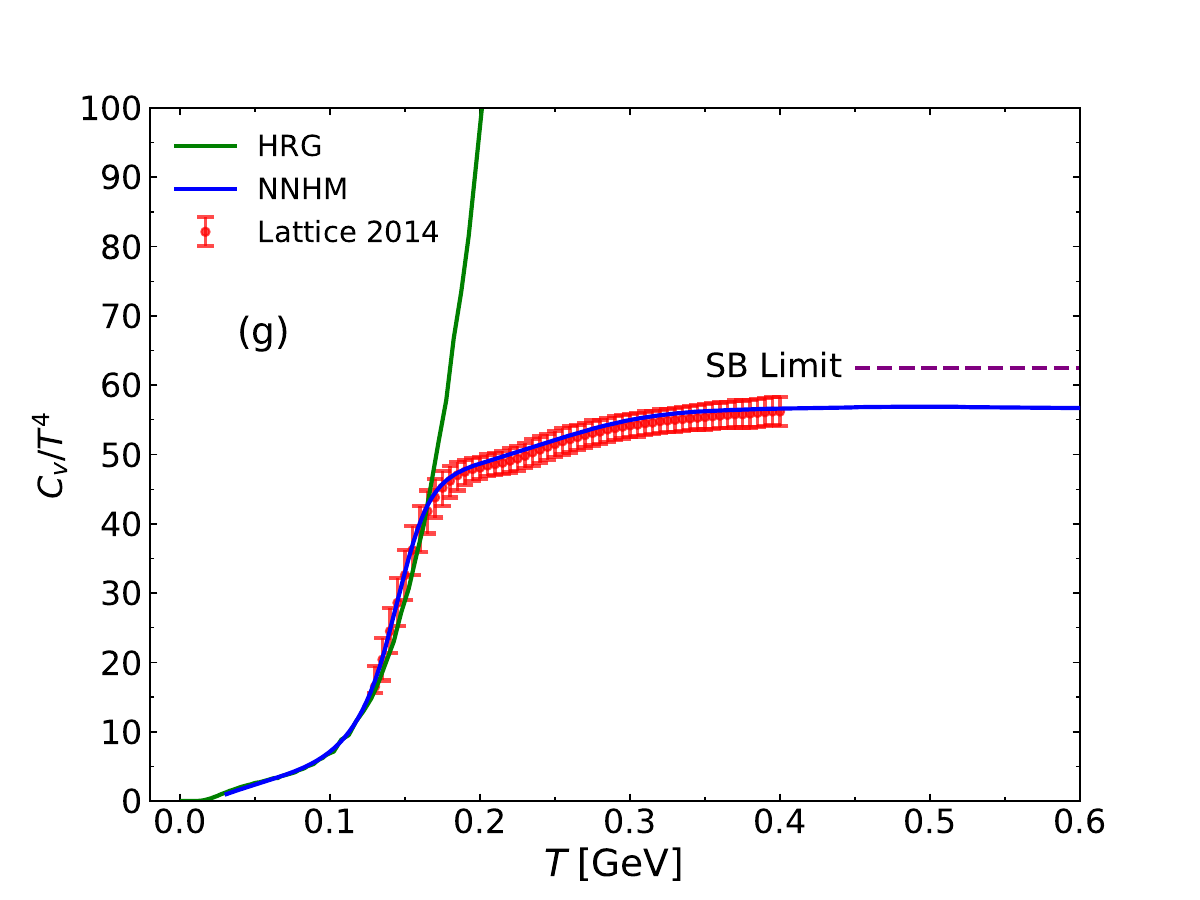}
    \end{minipage}%
    \hspace{0.1cm}%
    \begin{minipage}{0.45\textwidth}
        \centering
        \includegraphics[width=\textwidth]{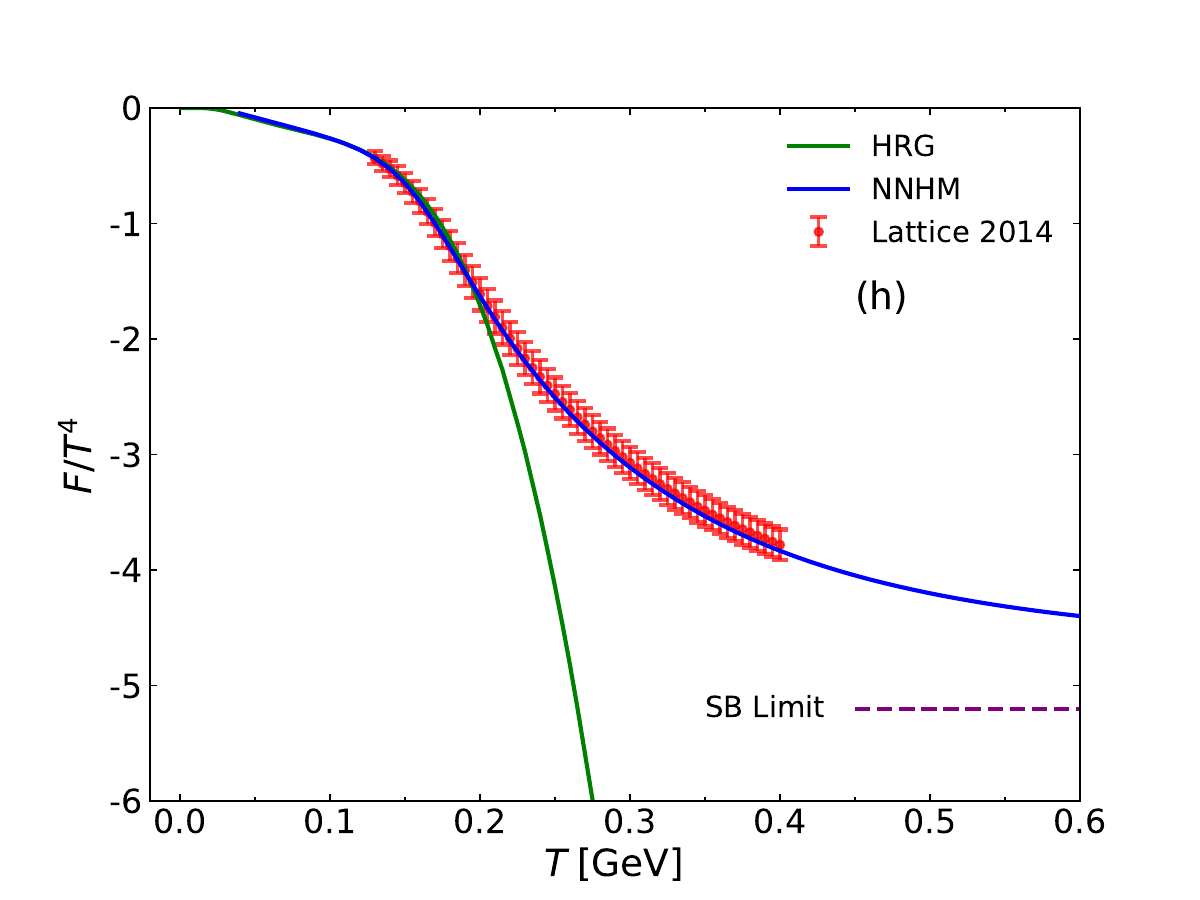}
    \end{minipage}
    \vspace{0.0cm} 
    \caption{Comparison of NNHM predictions with LQCD results for thermodynamic quantities at vanishing chemical potential. The blue solid lines represent the NNHM results, while the red points with error bars denote the (2+1)--flavor LQCD data from Ref.~\cite{HotQCD:2014kol, Bellwied:2015lba}. For reference, the green solid lines show the HRG model calculations, and the purple dashed lines indicate the SB limits. The plotted quantities are: entropy density $s$, energy density $\epsilon$, pressure $P$, trace anomaly $\epsilon - 3P$, second-order baryon susceptibility $\chi_2^B$, squared speed of sound $C_s^2$, specific heat $C_V$, and free energy $F$, all appropriately scaled by powers of temperature $T$.}
    \label{Fig.1}
\end{figure*}

At vanishing chemical potential, our machine-learning augmented EMD model demonstrates quantitative agreement with first-principles thermodynamics. 
As shown in Fig.\ref{Fig.1}, where we present the temperature dependence of the thermodynamic EoS at vanishing baryon chemical potential, the normalized pressure $P$, energy density $\epsilon$, and entropy density $s$ align closely with lattice benchmarks across the range $0.13 \text{ GeV} \leqslant T \leqslant 0.60 \text{ GeV}$. 

Crucially, the model captures the non-conformal behavior of the crossover transition. The speed of sound squared $C_s^2$ exhibits a characteristic dip near $T \approx 0.15$ GeV, representing the ``softest point" of the EoS.
While pure HRG models diverge significantly beyond this region as they fail to account for deconfined degrees of freedom, our model correctly below the Stefan--Boltzmann (SB) limit at high temperatures.
As shown in Fig.~\ref{Fig.1} , the NNHM prediction serves as a robust bridge between the low-temperature hadronic phase and the high-temperature QGP. 
The smooth interpolation from the hadronic dip to the conformal limit ($C_s^2 \to 1/3$) confirms that the neural network parameterization of $f(z)$ does not introduce spurious oscillations.
Furthermore, the second-order baryon susceptibility $\chi_2^B$ is well-reproduced, validating that the learned coupling $f(z)$ accurately reflects the system's sensitivity to density fluctuations.

\subsection{Finite Density Extrapolation and Susceptibility Response}

By extending the neural-network-determined $A(z)$ and $f(z)$ to finite chemical potential, we investigate the system's response to increasing baryon density for chemical potential-to-temperature ratios up to $\mu_B/T = 2.5$. 
We observe a monotonic enhancement in all extensive thermodynamic observables, consistent with the physical expectation of increased quark populations at higher chemical potentials.

\begin{figure*}[!htbp]
    \centering
    \begin{minipage}[b]{0.48\textwidth}  
        \centering
        \includegraphics[width=\textwidth]{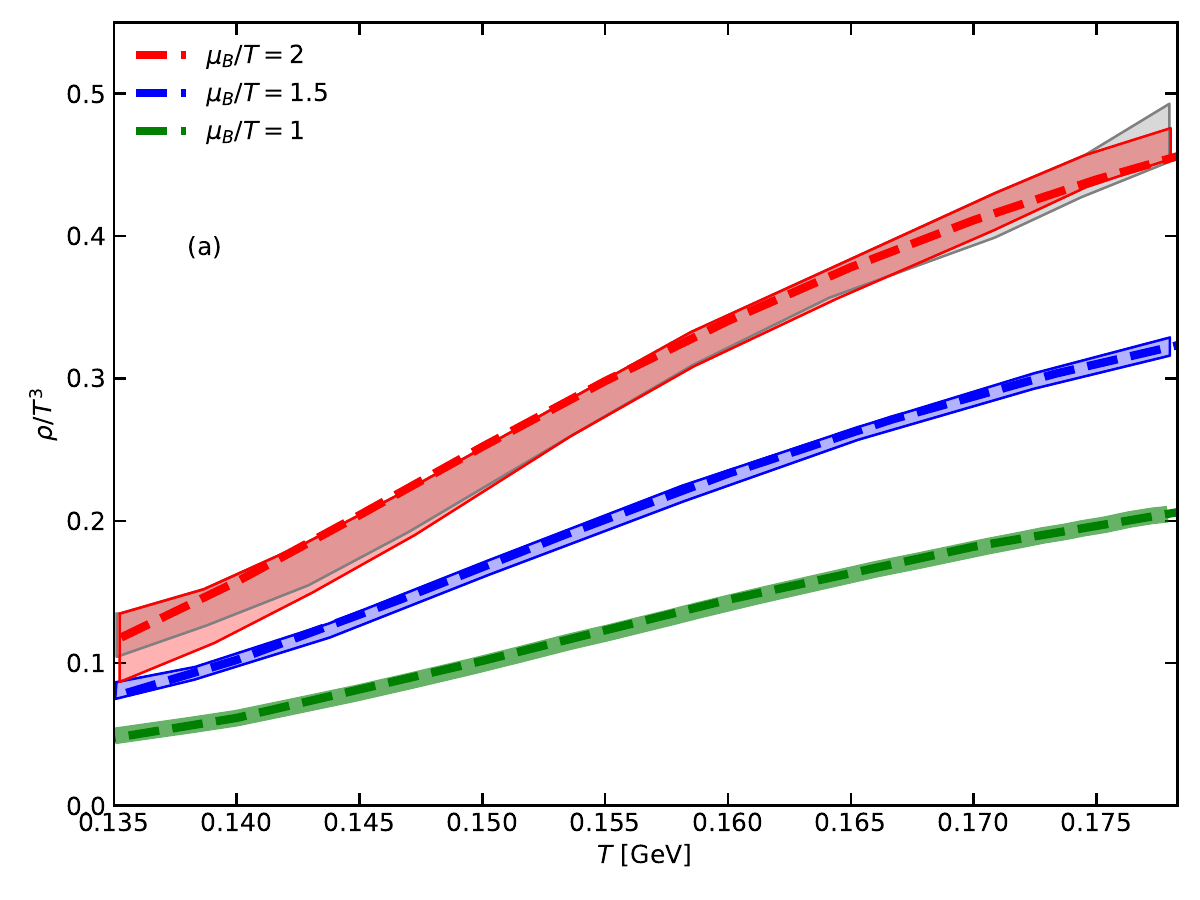}
        \vfill  
    \end{minipage}
    \begin{minipage}[b]{0.48\textwidth}
        \centering
        \includegraphics[width=\textwidth]{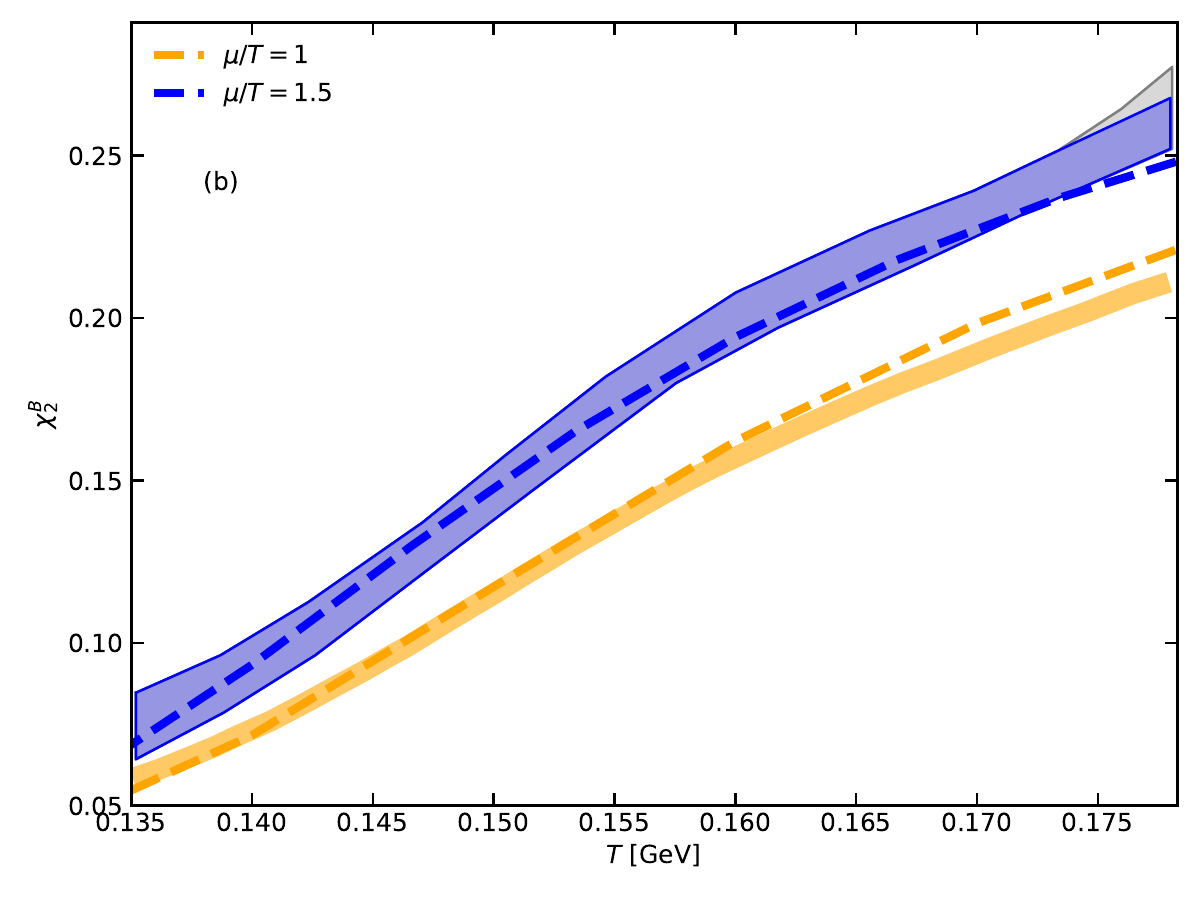}
        \vfill
    \end{minipage}
    
    \vspace{0.3cm}  
    
    \hspace*{0.25cm}%
    \begin{minipage}[b]{0.49\textwidth}
        \centering
        \includegraphics[width=\textwidth]{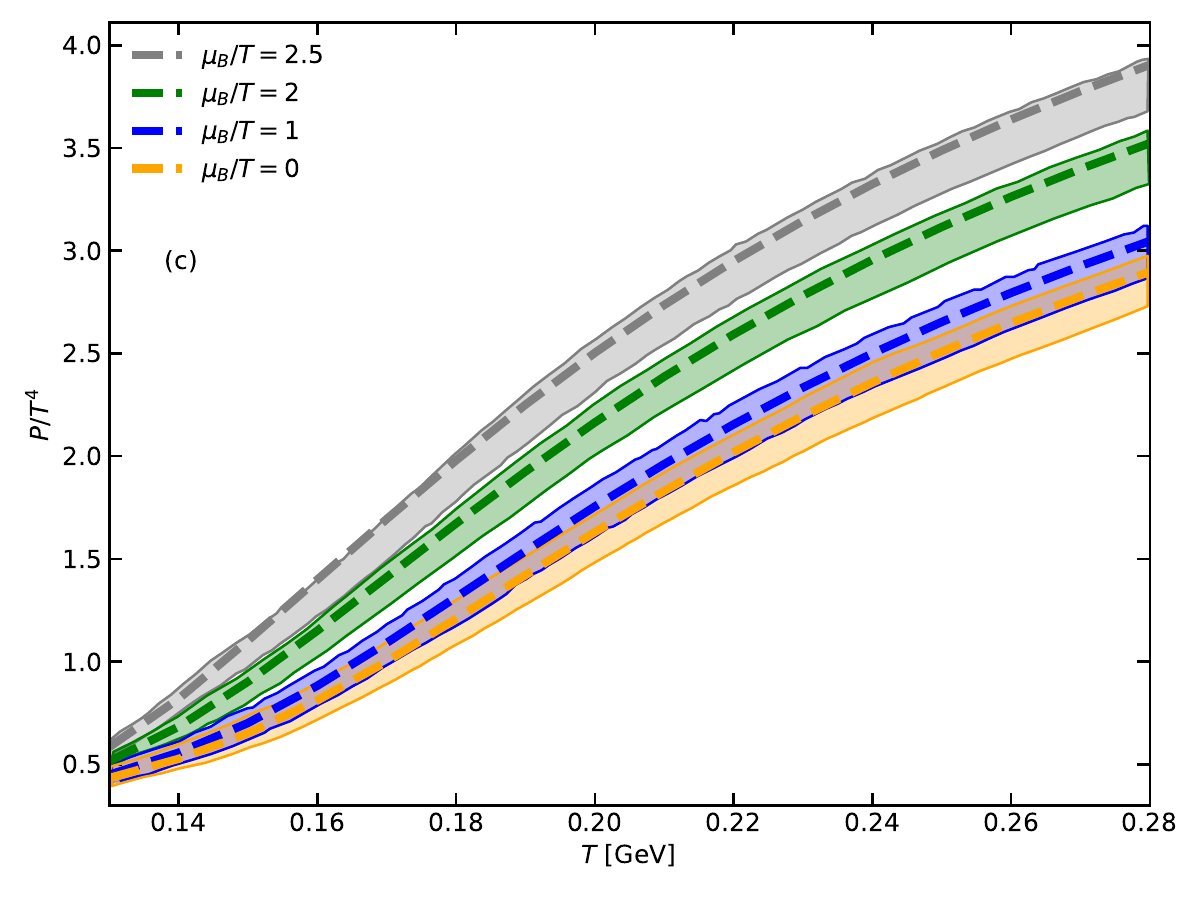}
        \vfill
    \end{minipage}%
    \begin{minipage}[b]{0.49\textwidth}
        \centering
        \includegraphics[width=\textwidth]{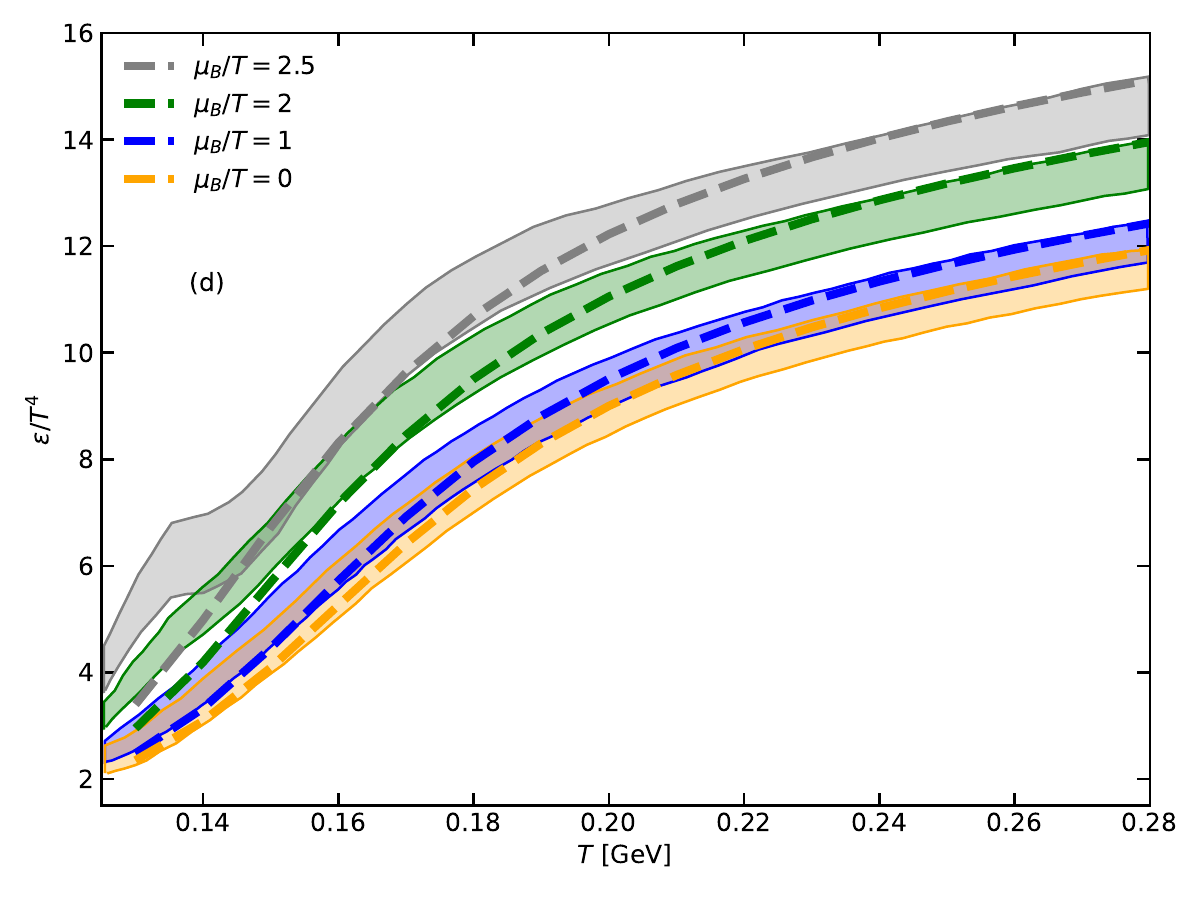}
        \vfill
    \end{minipage}
    \caption{Comparison of NNHM predictions with (2+1)-flavor LQCD results for the thermodynamic EoS at finite chemical potential, the shadow region represents the LQCD results, while the dashed line shows the calculations from NNHM. The panels display the temperature dependence of: (a) scaled baryon number density $\rho/T^3$, (b) second-order baryon susceptibility $\chi_2^B$, (c) scaled pressure $P/T^4$, and (d) scaled energy density $\epsilon/T^4$. The LQCD data are taken from Ref.~\cite{Bollweg:2022rps} (for $\rho$ and $\chi_2^B$) and Ref.~\cite{Bazavov:2017dus} (for $\epsilon$ and $P$).}
    \label{Fig.2}
\end{figure*}

The robustness of this extrapolation is cross-verified against lattice Taylor expansion and Padé approximant results at small $\mu_B$.
For ratios $\mu_B/T = 1.0$ and $1.5$, the holographic predictions for the scaled baryon density $\rho/T^3$ and susceptibility $\chi_2^B$ exhibit broad consistency with lattice estimates.
While minor deviations are observed in the magnitude of the baryon susceptibility and energy density at intermediate temperatures (as seen in Fig.~\ref{Fig.2} (b) and (d)), the model successfully captures the critical qualitative trends: the rapid rise of thermal quantities near $T_c$ and the correct asymptotic behavior.

\begin{figure*}[!htbp]
    \centering
    \includegraphics[width=0.9\linewidth]{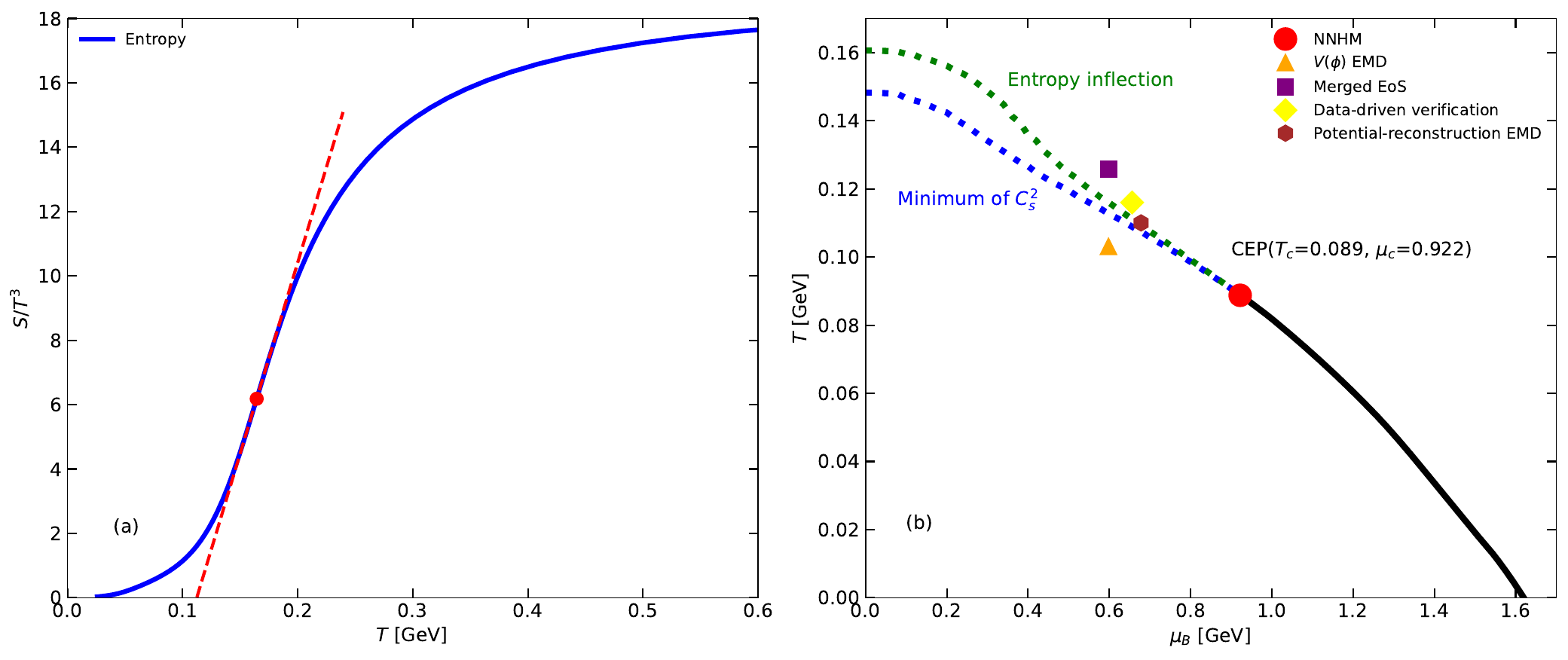}
\caption{QCD phase structure obtained within the data-driven NNHM. (a) Determination of the pseudo-critical temperature from the entropy density, where the inflection point of $s/T^3$ is identified through the tangent construction. (b) The QCD phase diagram in the $T$--$\mu_B$ plane. The crossover region is bounded by two pseudo-critical indicators: the minimum of the squared speed of sound ($C_s^2$, blue dotted line) and the entropy inflection line (green dotted line). These lines become indistinguishable at the CEP, located at $(T, \mu_B)=(0.089, 0.922)\mathrm{GeV}$ (red circle), beyond which a first-order phase transition line (solid black curve) emerges. For comparison, CEP estimates from the $V(\phi)$ EMD model (orange triangle)~\cite{Yang:2026brr}, the merged equation-of-state model including LQCD and HRG constraints (purple square)~\cite{Yang:2026brr}, and a potential-reconstruction EMD model (brown hexagon) are shown~\cite{Chen:2025goz}. The yellow diamond represents a synthetic-data benchmark, in which thermodynamic data generated from a known analytical EMD model are used to train the NNHM, demonstrating that the neural network can reproduce the corresponding CEP location within numerical uncertainties and confirming the data-driven nature of the reconstruction.}
    \label{Fig.3}
\end{figure*}

\subsection{Localization of the CEP and Phase Structure}

The central objective of this work is the precise mapping of the QCD phase structure across the $(T, \mu_B)$ plane. 
To delineate the phase boundaries, we track two distinct thermodynamic indicators: the local minimum of the squared speed of sound ($C_s^2$), which signifies the softening of the EoS, and the inflection point of the scaled entropy density ($s/T^3$), which traces the rapid liberation of degrees of freedom.

In the regime of low baryon chemical potential, these two pseudo-critical lines remain distinct, bounding a crossover region of finite width. 
As the baryon chemical potential increases, the separation between the pseudo-critical lines decreases. 
They approach each other and become indistinguishable near a singular point, signaling the termination of the crossover region and the emergence of a first-order phase transition line. 
Based on our optimized holographic reconstruction (as shown by the red dots in Fig. \ref{Fig.3}), we localize the CEP at:
\begin{equation}
T_c = 0.089 \text{ GeV}, \quad \mu_c = 0.922 \text{ GeV}.
\end{equation}

For baryon chemical potentials beyond the CEP ($\mu_B > \mu_c$), the system undergoes a first-order phase transition characterized by discontinuous jumps in entropy density and baryon density. 
The corresponding coexistence line extends toward the zero-temperature axis, terminating at a critical chemical potential of approximately $\mu_B \approx 1.6$ GeV at $T = 0$.

This topological structure is qualitatively consistent with predictions from functional approaches, including the Functional Renormalization Group (FRG)~\cite{Fu:2019hdw,Zhang:2016sje,Osman:2025,Roth:2024,Murgana:2024,Isserstedt:2019} and the Dyson-Schwinger Equations (DSE)~\cite{Gao:2016hno,Qin:2011dd,Shi:2014tka,Fischer:2013eca}. 
Compared with previous holographic studies, the CEP obtained in our model appears at a larger chemical potential due to the inclusion of HRG constraints. 
To illustrate this effect, several representative results are shown in Fig.~\ref{Fig.3}. 
The brown hexagon denotes a potential-reconstruction EMD model, which lies close to the $V(\phi)$ EMD model represented by the orange triangle~\cite{Yang:2026brr}. 
The purple square corresponds to the merged EoS model incorporating both LQCD and HRG data~\cite{Yang:2026brr}. 
These comparisons demonstrate that the inclusion of HRG data significantly shifts the CEP location toward larger baryon chemical potential within the holographic framework.

\subsection{Analytic Reconstruction via Symbolic Regression}

To interpret the numerical solutions generated by the neural networks, we employ symbolic regression (via PySR) to extract precise analytic forms for the warp factor $A(z)$ and gauge kinetic function $f(z)$. 
This process transforms the ``black-box'' output into transparent algebraic expressions.

The symbolic regression was implemented using genetic programming algorithms to search the functional space for expressions that minimize the mean squared error while maximizing simplicity. 
To strictly enforce holographic boundary conditions, we applied domain-specific pre-processing: for the warp factor, the regression target was transformed to $A(z)/z$ to guarantee the linear asymptotic behavior $A(z) \sim z$ near the AdS boundary ($z \to 0$). 
For the gauge kinetic term, we fitted the logarithm $\log(f(z)/f_0)$ to ensure the function remains positive definite. 
Furthermore, we imposed a custom weighting scheme, $w(z)$, which heavily penalizes errors in the UV region ($z < 0.2$) to preserve the accuracy of the perturbative slope. 
The search space was restricted to fundamental arithmetic operators $\{+, -, \times, \div, \text{square}\}$ and excluded general power laws, thereby forcing the algorithm to find analytic rational or exponential structures rather than unphysical fractional approximations.

\begin{figure*}[!htbp]
    \centering
    \includegraphics[width=0.9\linewidth]{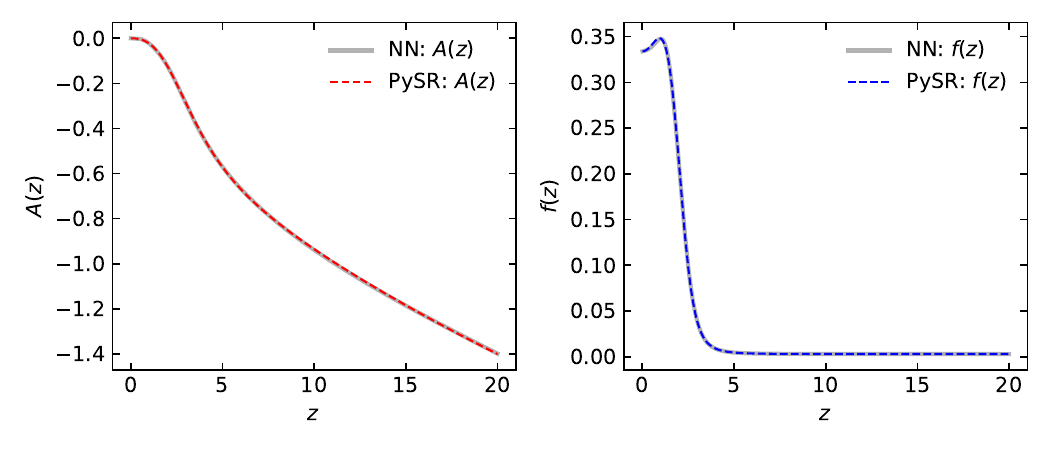}
    \caption{Comparison of the neural-network numerical solutions (gray solid lines) with the analytical expressions obtained from symbolic regression (colored dashed lines). The left panel displays the warp factor $A(z)$, while the right panel shows the gauge kinetic function $f(z)$. The nearly perfect overlap ($R^2 > 0.999$) demonstrates that the derived analytical formulas capture the essential nonlinear dynamics and pole structures of the holographic model.}
    \label{Fig.4}
\end{figure*}

The resulting analytic expressions achieve remarkable consistency with the numerical data ($R^2 > 0.999$). 
Crucially, these forms explicitly reveal the pole structures in the bulk geometry that drive the confinement-deconfinement transition. The reconstructed functions are:
\begin{equation}
A(z) = z \left[ \left(a + \frac{b}{c - \left(z + \frac{d}{z}\right)}\right) (z + e) \right],
\end{equation}

\begin{equation}
f(z) = f_0 \exp\left[ f + \frac{g}{h \left(i + \frac{1}{j - z^{2}}\right)^{2} + k} \right].
\end{equation}
The corresponding parameters are listed in Table~\ref{Table.1}, and the comparison with the numerical results is shown in Fig.~\ref{Fig.4}.

\begin{table}[htbp]
\centering
\small
\setlength{\tabcolsep}{12pt}
\caption{Analytical parameters for the warp factor $A(z)$ and the gauge kinetic function $f(z)$ reconstructed via symbolic regression. The functional forms are constrained to rational and exponential structures to satisfy holographic boundary conditions. Parameters $a$--$e$ define the rational structure of $A(z)$, while $f_0$ and $f$--$k$ characterize the normalization and near-horizon behavior of $f(z)$.}
\begin{tabular}{@{}cc@{\hspace{24pt}}cc@{}}
\toprule[1pt]
\textbf{Parameter} & \textbf{Value} & \textbf{Parameter} & \textbf{Value} \\
\midrule[0.5pt]
$a$   & $4.83\times10^{-4}$ & $f$ & $0.0403$  \\
$b$   & $6.16\times10^{-2}$ & $g$ & $7.40$    \\
$c$   & $3.30$              & $h$ & $-134$    \\
$d$   & $11.7$              & $i$ & $0.00743$ \\
$e$   & $2.74$              & $j$ & $0.885$   \\
$f_0$ & $0.33448$           & $k$ & $-1.55$   \\
\bottomrule[1pt]
\end{tabular}
\label{Table.1}
\end{table}

\subsection{Three-Dimensional Structure of the QCD Equation of State}
\begin{figure*}[htbp]
    \centering
    \begin{minipage}{0.45\textwidth}
        \centering
        \includegraphics[width=\textwidth]{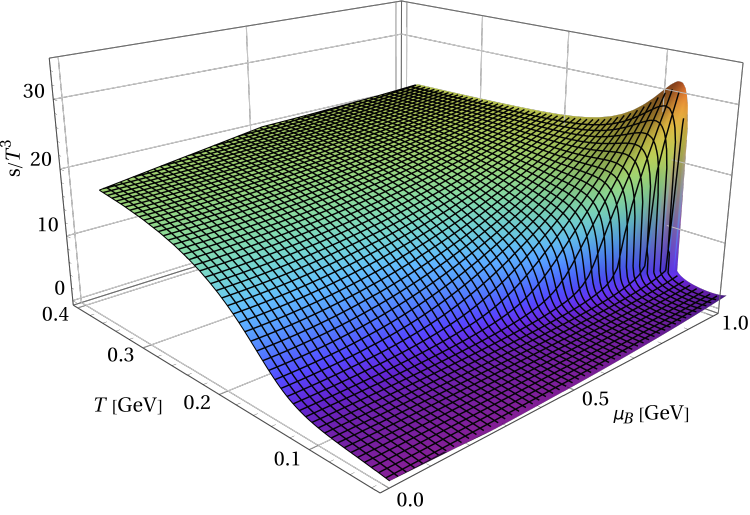}
    \end{minipage}
    \hspace{0.1cm}
    \begin{minipage}{0.45\textwidth}
        \centering
        \includegraphics[width=\textwidth]{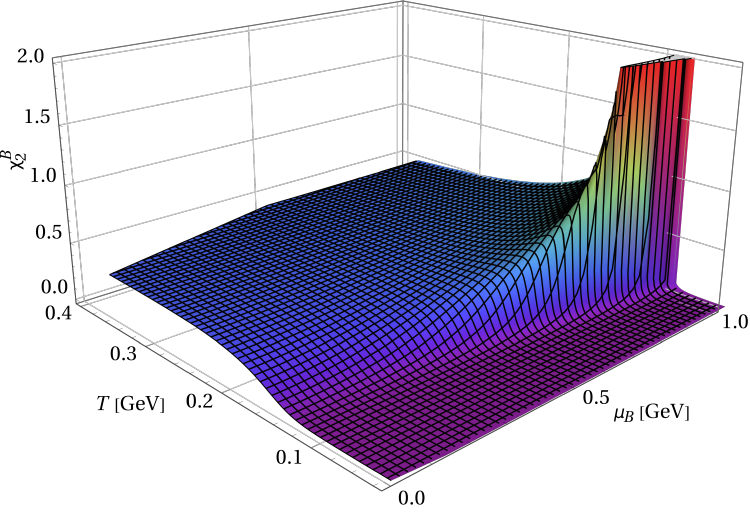}
    \end{minipage}
    \vspace{0.0cm}
    \begin{minipage}{0.45\textwidth}
        \centering
        \includegraphics[width=\textwidth]{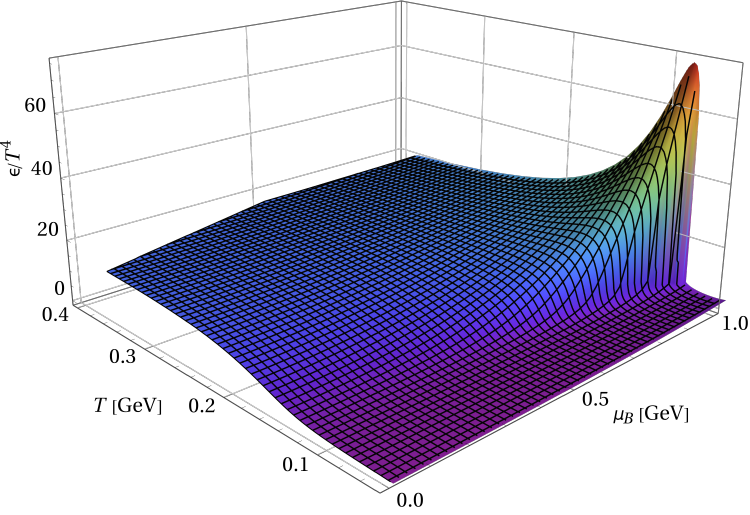}
    \end{minipage}%
    \hspace{0.1cm}%
    \begin{minipage}{0.45\textwidth}
        \centering
        \includegraphics[width=\textwidth]{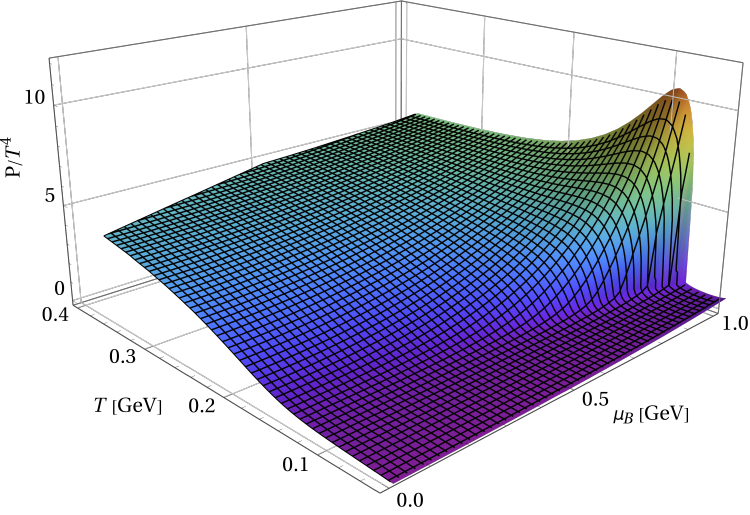}
    \end{minipage}
    \vspace{0.0cm} 
    \begin{minipage}{0.45\textwidth}
        \centering
        \includegraphics[width=\textwidth]{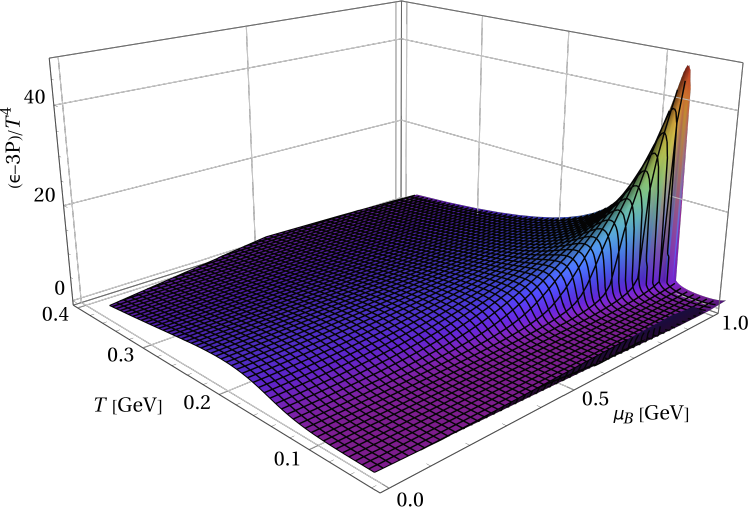}
    \end{minipage}
    \hspace{0.1cm}
    \begin{minipage}{0.45\textwidth}
        \centering
        \includegraphics[width=\textwidth]{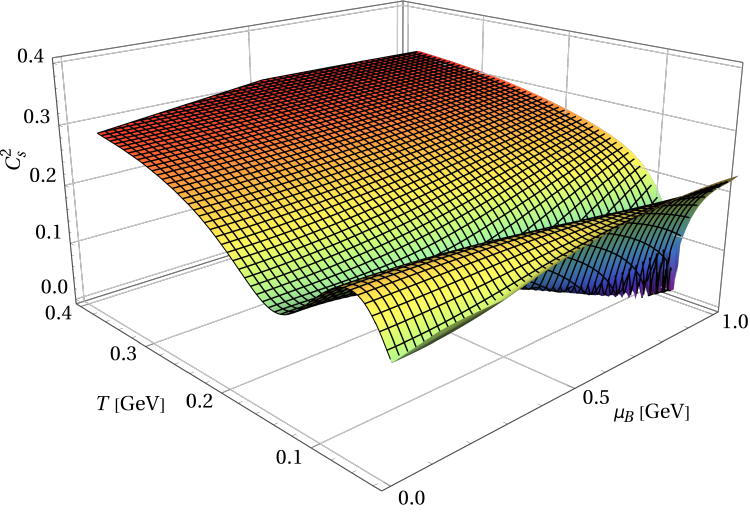}
    \end{minipage}
    \vspace{0.0cm}
    \begin{minipage}{0.45\textwidth}
        \centering
        \includegraphics[width=\textwidth]{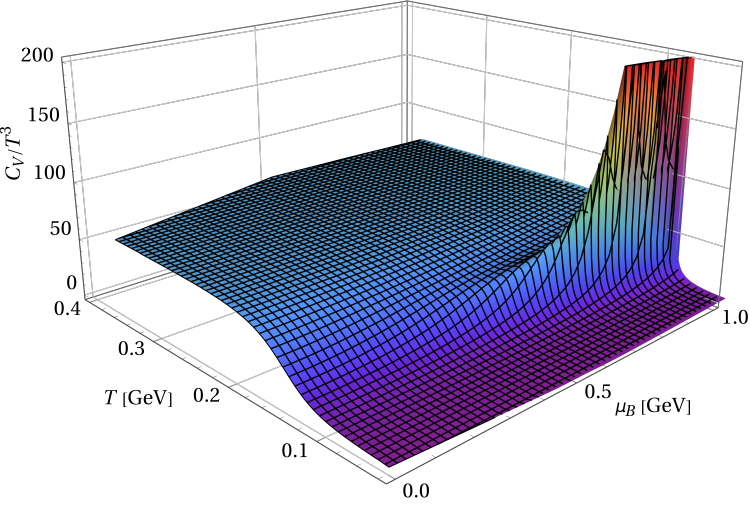}
    \end{minipage}%
    \hspace{0.1cm}%
    \begin{minipage}{0.45\textwidth}
        \centering
        \includegraphics[width=\textwidth]{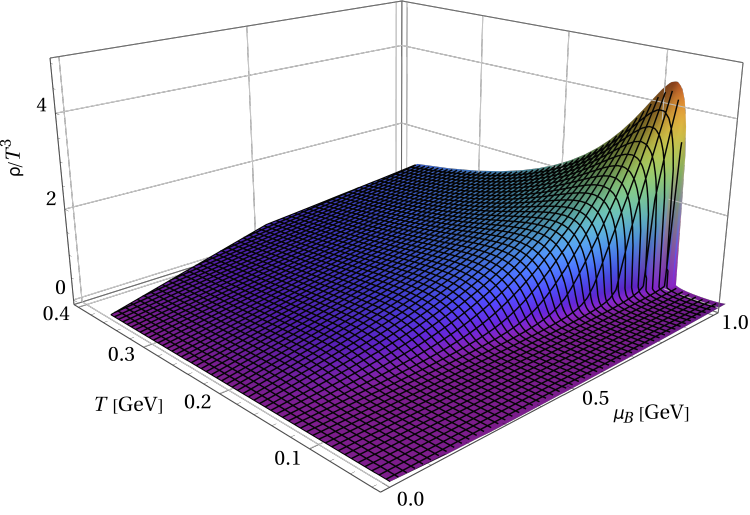}
    \end{minipage}
    \vspace{0.2cm} 
    \caption{Three-dimensional distributions of thermodynamic observables obtained from the reconstructed neural-network holographic model on the $(T,\mu_B)$ plane. The panels show, from left to right and top to bottom, the entropy density $s/T^3$, the second-order baryon number susceptibility $\chi_2^B$, the energy density $\epsilon/T^4$, the pressure $p/T^4$, the trace anomaly $(\epsilon-3p)/T^4$, the squared speed of sound $C_s^2$, the specific heat $C_V/T^3$, and the baryon number density $\rho/T^3$. These surface plots illustrate the global thermodynamic structure of the model at finite temperature and baryon chemical potential, and provide a direct visualization of the nontrivial density dependence.}
    \label{Fig.5}
\end{figure*}

Based on the analytically reconstructed warp factor $A(z)$ and gauge kinetic function $f(z)$, we evaluate the thermodynamic observables over the full $(T,\mu_B)$ plane. 
The corresponding three-dimensional profiles are displayed in Fig.~\ref{Fig.5}, including the entropy density, baryon number susceptibility, energy density, pressure, trace anomaly, squared speed of sound, specific heat, and baryon density. 
These surfaces provide a global view of the EoS at finite temperature and finite baryon chemical potential.

As shown in Fig.~\ref{Fig.5}, the extensive thermodynamic quantities, including $s/T^3$, $\epsilon/T^4$, $P/T^4$, $C_V/T^3$, and $\rho/T^3$, generally increase with increasing chemical potential, especially in the low-temperature region. This behavior reflects the enhanced population of baryonic degrees of freedom at finite density.

The trace anomaly $(\epsilon-3P)/T^4$ develops a pronounced structure in the low-$T$, high-$\mu_B$ region, indicating stronger non-conformal dynamics there. 
In contrast, the squared speed of sound $C_s^2$ exhibits a nontrivial suppression around the crossover region and becomes softer as the system approaches the phase boundary, which is consistent with the expected softening of the EoS near criticality.

The baryon number susceptibility $\chi_2^B$ and the baryon density $\rho/T^3$ show particularly strong responses to finite chemical potential, demonstrating that the reconstructed gauge sector captures the density dependence of the system in a thermodynamically consistent manner. 
Overall, the smooth but nontrivial structures of these surfaces support the reliability of the neural-network holographic reconstruction in describing QCD thermodynamics across a broad region of the phase diagram.

\subsection{Data-driven verification}

To explicitly demonstrate that the neural network does not introduce spurious bias, we perform a benchmark reconstruction using synthetic data.
A natural concern in machine-learning-assisted holography is whether the predicted CEP location (or even its existence) could be an artifact of the chosen neural architecture rather than being dictated by the input thermodynamics.
To address this, we feed the same NNHM training pipeline with synthetic thermodynamic data—specifically the entropy density $s$, the second-order baryon susceptibility $\chi_2^B$, and the baryon density $\rho$ generated from the analytical EMD model used in our previous work~\cite{Chen:2025goz}.
As indicated by the yellow diamond in Fig.~\ref{Fig.3}, the neural network accurately reproduces the CEP of that reference model.
This consistency check confirms that the NNHM acts as an unbiased functional mapper: the trajectories of the pseudo-critical lines and their convergence at the CEP are dictated by the physics encoded in the input data rather than by restrictive parametric assumptions.

Quantitatively, the CEP (yellow diamond) extracted from data generated by the potential-reconstruction EMD model is located at $(T_c, \mu_c)=(0.116, 0.656) \mathrm{GeV}$, which is close to the value obtained in the original model, $(T_c, \mu_c)=(0.110, 0.678) \mathrm{GeV}$, reported in Ref.~\cite{Chen:2025goz}, within the expected numerical uncertainties.
This agreement supports the reliability of our reconstruction pipeline and indicates that the CEP prediction of the NNHM is genuinely data-driven.


\section{Summary} \label{sec.IV}

In this work, we developed a neural-network-based holographic reconstruction of the QCD EoS across the $(T,\mu_B)$ plane by combining high-temperature LQCD inputs with low-temperature constraints from the Hadron Resonance Gas (HRG) model. 
Within a bottom-up Einstein–Maxwell–Dilaton (EMD) framework, the metric warp factor $A(z)$ and the gauge kinetic function $f(z)$ were parameterized by neural networks and optimized using thermodynamic observables, providing a flexible and data-driven solution to the holographic inverse problem without relying on restrictive analytic ansätze.

At vanishing baryon chemical potential, the reconstructed model reproduces the lattice EoS and captures the non-conformal features of the QCD crossover, while remaining consistent with HRG thermodynamics in the confined phase. 
Extending the analysis to finite baryon chemical potential, we obtained a self-consistent set of thermodynamic observables and mapped the QCD phase structure using the minima of the squared speed of sound $C_s^2$ and the inflection points of the scaled entropy density $s/T^3$ as pseudo-critical indicators. 
The resulting phase diagram exhibits a finite-width crossover region at small $\mu_B$, which terminates at a critical endpoint located at $(T_c,\mu_c)=(0.089,0.922)\,\mathrm{GeV}$, beyond which a first-order transition line emerges.

To visualize the global thermodynamic structure of the reconstructed model, we further constructed three-dimensional surfaces of the EoS across the $(T,\mu_B)$ plane, including the entropy density, baryon density, pressure, energy density, trace anomaly, speed of sound, and specific heat. 
These surfaces provide a comprehensive view of the density dependence of QCD thermodynamics and illustrate the softening behavior of the EoS in the vicinity of the phase-transition region.

To improve the interpretability of the neural-network reconstruction, we applied symbolic regression to extract compact analytic expressions for the background functions $A(z)$ and $f(z)$, achieving excellent agreement with the numerical solutions. 
Finally, a synthetic-data benchmark based on an analytical EMD model demonstrates that the reconstruction pipeline can faithfully recover the corresponding critical endpoint, confirming that the predicted phase structure is primarily determined by the input thermodynamic data rather than by model assumptions.

However, several limitations must be acknowledged. First, the reconstruction's fidelity is inherently tied to the quality and scope of the input data. While we used state-of-the-art lattice results at $\mu_B=0$, the constraints at finite $\mu_B$ are indirect, derived from the neural network's interpolation/extrapolation based on the zero-density training and the HRG boundary condition. 
The absence of direct, precise lattice data at finite $\mu_B$ remains a fundamental challenge. 
Second, the EMD framework, while versatile, incorporates a specific set of fields (gravity, a $U(1)$ gauge field for baryon number, and a dilaton for running coupling). 
It does not explicitly include chiral symmetry breaking, which is a key aspect of the QCD transition. 
The phase structure we identify is therefore primarily associated with deconfinement and thermodynamic crossover/transition, as probed by bulk observables like the speed of sound. 
Third, the symbolic regression, while successful in providing interpretable analytic forms for $A(z)$ and $f(z)$, yields approximations. 
The excellent agreement with numerical results is encouraging, but the expressions may not capture all nuances in regions with sparse data constraints.

In conclusion, we have established a powerful, flexible, and data-driven pipeline for holographic QCD thermodynamics. 
By leveraging neural networks to solve the inverse problem, we obtained a comprehensive model that reproduces known lattice results, predicts a phase structure featuring a critical endpoint, and provides global thermodynamic surfaces. 
The synergy between holography and machine learning demonstrated here offers a compelling paradigm for extracting the gravitational dual of complex quantum field theories from empirical and ab initio data, paving the way for a more rigorous and data-informed application of gauge/gravity duality to strong interaction physics.

Overall, this work highlights the potential of machine-learning-assisted holographic approaches as a data-driven framework for exploring the QCD phase diagram. 
Future developments may incorporate more precise lattice data and additional QCD observables, which could further constrain the holographic background and lead to a more reliable determination of the thermodynamic properties of strongly interacting matter.

\acknowledgments

We would like to thank Yan-Qing Zhao and Jia-Hao Wang for helpful discussions. This work is supported in part National Key Research and Development Program of China under Contract No. 2022YFA1604900 and by the National Natural Science Foundation of China (NSFC) Grant No. 12405154, No.92570117, No.12435009 and No. 12275104, Shenzhen Peacock fund under No. 2023TC0179, Ministry of Science and Technology of China under Grant No. 2024YFA1611004, the European Union -- Next Generation EU through the research grant number P2022Z4P4B ``SOPHYA - Sustainable Optimised PHYsics Algorithms: fundamental physics to build an advanced society'' under the program PRIN 2022 PNRR of the Italian Ministero dell'Universit\`a e Ricerca (MUR).


\bibliographystyle{JHEP}
\bibliography{biblio.bib}

\end{document}